\documentclass[prc,twocolumn,showpacs,preprintnumbers,amsmath,amssymb]{revtex4-1}
\usepackage[utf8]{inputenc}
\usepackage{graphicx}

\allowdisplaybreaks

\begin{document}
\title{Photo- and electroproduction of $K^{+}\Lambda$ with unitarity-restored isobar model}
\author{D. Skoupil$^{1,2}$ and P. Byd\v{z}ovsk\'y$^{1}$} 
\affiliation{$^{1}$Nuclear Physics Institute, CAS, \v{R}e\v{z}/Prague, 
Czech Republic,\\
$^{2}$Nishina Center for Accelerator-Based Science, RIKEN, Wako, 351-0198, Japan}

\date{\today }
 
\begin{abstract}
Exploiting the isobar model, kaon photo- and electroproduction on the proton in the resonance region comes under scrutiny. An upgrade of our previous model, comprising higher-spin nucleon and hyperon exchanges in the consistent formalism, was accomplished by implementing energy-dependent widths of nucleon resonances, which leads to a different choice of hadron form factor with much softer values of cutoff parameter for the resonant part. For a reliable description of electroproduction, the necessity of including longitudinal couplings of nucleon resonances to virtual photons was revealed. We present a new model whose free parameters were adjusted to photo- and electroproduction data and which provides a reliable overall description of experimental data in all kinematic regions. The majority of nucleon resonances chosen in this analysis coincide with those selected in our previous analysis and also in the Bayesian analysis with the Regge-plus-resonance model as the states contributing to this process with the highest probability.
\end{abstract}


\maketitle

\section{Introduction}
Main goals of the investigation of kaon photo- and electroproduction from nucleon target are the study of baryon resonance spectrum and interactions in systems of hyperons and mesons. It can also provide information about the existence and characteristics of ``missing" resonances which have been predicted by quark models~\cite{Capstick,Loring}. On top of that, the right description of the elementary production process is essential for obtaining dependable predictions of the cross sections in hypernucleus electroproduction.

There are various ways of describing the elementary photo- and electroproduction process. One of the methods is based on effective Lagrangians with hadron degrees of freedom only. As above the threshold of kaon production there are other channels already open (\emph{e.g.}, $\pi N$ or $\eta N$) and they couple to the kaon-production channel by means of meson-baryon interaction, one should take all these channels into account so as not to violate unitarity. This is done in the coupled-channels models~\cite{cc}; a shortcoming of such an approach is that there is a lack of information on some processes, \emph{e.g.}, $K^+\Lambda\rightarrow K^+\Lambda$. This can be, however, avoided by neglecting the coupled-channels effects and assuming that their contributions are, to some extent, comprised by means of effective values of the coupling constants adjusted to experimental data. This assumption was adopted in a number of single-channel isobar models, \emph{e.g.}, Kaon-MAID (KM), Saclay-Lyon (SL), which are used for description of the production process and also in the calculations of the electroproduction of hypernuclei. The unitarity corrections in these single-channel models can be included by energy-dependent widths of nucleon resonances in the resonance propagators~\cite{unitarity}.

Recently, we have proposed a variant of the phenomenological single-channel isobar model for kaon photoproduction on the nucleon, namely the $p(\gamma,K^+)\Lambda$ process~\cite{BS}. Although this model was constructed for photoproduction, its extension towards electroproduction can be done in a straightforward way as dependence on the photon mass was kept in the formalism. In constructing this model, we made use of the consistent formalism for the description of high-spin baryon resonances~\cite{Pascalutsa,Vrancx}, which allowed us to include among others also the spin-3/2 hyperon resonances in the \emph{u} channel. In the current publication we present and comment on the changes that were carried out on this model since the publication of Ref.~\cite{BS}: the insertion of energy-dependent decay widths of nucleon resonances, the introduction of electromagnetic form factors, and adding  new couplings of the photon longitudinal mode to nucleon resonance fields (longitudinal couplings).

This article is organized as follows: in Sec.~\ref{sec:IM} we present our model and the inclusion of its novel features - energy-dependent decay widths of $N^*$'s, electromagnetic form factors, and longitudinal couplings of virtual photons to nucleon resonances. The discussion of achieved outcomes and conclusions are presented in Secs.~\ref{sec:dis} and \ref{sec:con}, respectively.

\section{Single-channel isobar model}
\label{sec:IM}

In this section, we provide the reader with the main features of our approach. More details and formulae can be found in Ref.~\cite{BS}. We investigate the kaon-hyperon photoproduction on the proton in the energy ranging from the threshold of 1.609~GeV up to approximately 2.5~GeV. In the isobar model, the amplitude is constructed using effective meson-baryon Lagrangian as a sum of tree-level Feynman diagrams representing the \emph{s-}, \emph{t-}, and \emph{u-}channel exchanges of ground state hadrons and their excited states. In this approach we neglect the higher-order contributions which account for, \emph{e.g.}, the rescattering effects. From the multiplicity of Feynman diagrams contributing to the amplitude, only the exchanges of nucleon resonances in the \emph{s} channel can make a resonant structure in the observables; the remaining diagrams create the background since the poles of corresponding exchanged resonances are far from the physical region.

The kaon photoproduction occurs in the third nucleon-resonance region where a huge number of nucleon resonances exist. There is therefore no dominant $N^*$ in the kaon photoproduction, unlike in $\eta$ or $\pi$ photoproduction, and one has to consider \emph{a priori} more than 20 resonances with the mass $\leq 2\,\text{GeV}$. This then results in abundance of possible resonance configurations that should be examined, which leads to a plethora of models describing the data reasonably. 

The contributions of higher-spin nucleon and hyperon resonances are of paramount importance in the energy region under study. It is, however, known that a formalism for baryon fields with a higher spin is problematic due to presence of nonphysical degrees of freedom which are connected with the lower-spin modes of the Rarita-Schwinger (RS) field. If the RS field is off-shell, the nonphysical components can participate in the interaction, which we then call inconsistent. In our model, we make use of a consistent formalism~\cite{Pascalutsa,Vrancx} to secure that only the physical degrees of freedom contribute. The consistency is ensured by the requirement of the interaction Lagrangians being invariant under the local U(1) gauge transformation of the RS field. The fulfilment of this requirement then makes the corresponding vertices transversal to the momentum of the exchanged particle, which provides contributions from the highest-spin components of the RS off-mass-shell propagator only~\cite{BS}.

Another important feature of the RS gauge invariance is that the contributions of resonances with high spin are regular. This is particularly important for the exchanges of spin-3/2 hyperon resonances. In the past, mainly spin-1/2 hyperon resonances were introduced in the models; the only model assuming spin-3/2 hyperon resonances, although in the inconsistent formalism, is Saclay-Lyon C model~\cite{Mizutani}. The reason for this is that in kaon production the Mandelstam variable \emph{u} can be zero in the physical region, which makes the inclusion of spin-3/2 $Y^*$'s in the inconsistent formalism problematic due to a divergence in the spin-1/2 components of the RS propagator. In the consistent formalism, however, these components vanish in the amplitude and we can, therefore, include the spin-3/2 hyperon resonances together with spin-3/2 and spin-5/2 nucleon resonances. Detailed treatise on various resonances contributing to the process and the introduction of resonances with higher spin is given in Ref.~\cite{BS}.

A characteristic and well-known feature of the $p(\gamma,K^+)\Lambda$ process described by the isobar model is too large a contribution of Born terms to the cross section, which overshoots the data. The nonphysically large strength of Born terms have to be reduced so as to get a reliable description of the cross section and other observables, which then allows an analysis of the resonant content of the amplitude. This can be achieved either by inserting hadron form factors in the strong vertices or by assuming exchanges of $Y^*$'s in the \emph{u}-channel, which interfere destructively with the Born terms. We combine both approaches in our model. Needless to say, the inclusion of hadron form factors modifies values of coupling constants of the resonances and, therefore, the choice of the method for suppressing the Born-term contribution affects the dynamics of the model.

Not only do we introduce hadron form factors to suppress Born terms, we also need them to refine the behaviour of high-spin resonances whose contributions grow substantially with energy due to the high-power momentum dependence. Except for these suppression issues, the hadron form factor is introduced as it mimics the internal structure of hadrons in strong vertices, which is neglected in the hadrodynamical approach. There are many shapes of the hadron form factor at hand and one can opt for dipole, multidipole, Gaussian or multidipole Gaussian hadron form factor (see Ref.~\cite{BS} for their definitions). The efficacy of the form factor is strongly dependent on the value of its cutoff parameters, allowing unphysical behaviour to develop with higher cutoff values. The multidipole Gaussian form factor is the only one which is almost independent of the cutoff value.

The total amplitude of the isobar model is gauge invariant: contributions of the \emph{u-}channel Born term and all non Born terms are gauge invariant separately; the only gauge non invariant terms occurring in the \emph{s-} and \emph{t-}channel cancel one another in the sum of these contributions. However, when we introduce the hadron form factors, these gauge non-invariant terms no longer cancel and one is forced to insert a contact term to restore the gauge invariance.
 
In order to ensure regularity of the tree-level amplitude, we shift the poles corresponding to resonant states into the complex plane, $m_R\rightarrow m_R-i\Gamma_R/2$, introducing a decay width $\Gamma_R$ which takes into account the finite lifetime of the resonance. The Feynman propagator then acquires a form of
\begin{equation}
\frac{1}{\not\! q-m_R+{i}\Gamma_R /2} = 
\frac{\not\! q+m_R-{i}\Gamma_R /2}{q^2-m^2_R+{i}m_R\Gamma_R+\Gamma_R^{\,2} /4}
\label{eq:Fprop}
\end{equation}
and we assume the following approximation
\begin{equation}
\frac{\not\! q +m_R}{q^2-m_R^2 + {i}\,m_R\Gamma_R},
\label{eq:Fprop-approx}
\end{equation}
which is exploited, \emph{e.g.}, in Saclay-Lyon or Ghent isobar models. In the majority of isobar models the resonance widths are considered as constant parameters and the Breit-Wigner values from the Particle Data Tables are taken.

\subsection{Energy-dependent decay widths of the $N^*$'s}
In order to approximately account for the unitarity corrections at tree level, the energy-dependent decay widths of the nucleon resonances were used in the KM model. The energy dependence of the width $\Gamma_R$ is given by the possibility of a resonance to decay into various channels that are open. We assume the energy dependence of the decay widths of the form
\begin{equation}
\Gamma_R(s)=\Gamma_{N^*}\frac{\sqrt{s}}{m_{N^*}}\sum_i\left[x_i \left(\frac{|\vec{q}_i|}{|\vec{q}_i^{N^*}|}\right)^{2l+1}\!\frac{D_l(|\vec{q}_i|)}{D_l(|\vec{q}_i^{\,N^*}|)}\right],
\label{eq:Gamma-s}
\end{equation}
where $s=q^2$, $m_{N^*}$ is the resonance mass, and the sum goes over all possible decay channels into a meson and a baryon with masses $m_i$ and $m_b$, respectively, and with the relative orbital momentum $l$. Furthermore, $\Gamma_{N^*}$ and $x_i$ denote the total decay width, and a relative branching ratio of given resonance into the $i$-th channel (see Tab.~\ref{tab:BR}), respectively. Final-state momenta read
\begin{subequations}
\begin{align}
|\vec{q}_i^{N^*}|=& \sqrt{\frac{(m_{N^*}^2-m_b^2+m_i^2)^2}{4m_{N^*}}-m_i^2},\\
|\vec{q}_i|=& \sqrt{\frac{(s-m_b^2+m_i^2)^2}{4s}-m_i^2}
\end{align}
\end{subequations}
and the so-called fission barrier factor is
\begin{equation}
D_l(x) = \texttt{exp}\left(-\frac{x^2}{3\alpha^2}\right),
\end{equation}
with $\alpha=410\,\mbox{MeV}$ and $x= |\vec{q}_i|$ or $|\vec{q}_i^{\,N^*}|$. For $\sqrt{s}=m_{N^*}$, the decay width (\ref{eq:Gamma-s}) reduces to
\begin{equation}
\Gamma_R(m_{N^*}^2)=\Gamma_{N^*}\sum_i x_i,
\end{equation}
and for $\sum_i x_i =1$ it holds $\Gamma_R(m_{N^*}^2)=\Gamma_{N^*}$.

Dependence of the width on the relative momentum $l$ is important in the energy region 
of our interest as for high-spin resonances the factor $|\vec{q}_i|/|\vec{q}_i^{N^*}|$ 
outweighs the exponential factor $D_l(|\vec{q}_i|)/D_l(|\vec{q}_i^{N^*}|)$ and therefore it dominates the energy dependence of the width.
The Feynman propagator for the nucleon resonances is then used in the form
\begin{equation}
\frac{\not\! q +m_{N^*}}{q^2-m_{N^*}^2 + i\,m_{N^*}\Gamma_R(s)}.
\end{equation}
The introduction of energy-dependent widths for nucleon resonances affects also the choice of hadron form factor and its cutoff parameters, which will be discussed later on in Sec.~\ref{sec:dis}. The branching ratios $x_i$ of the nucleon resonances used in our extended model are listed in Table~\ref{tab:BR}.

\begin{table}[h]
\begin{center}
\begin{tabular}{l r r r r}
\hline \hline
                  & $N\pi$      & $N\pi\pi$ & $N\eta$     & $K\Lambda$  \\ \hline
N1 $P_{11}(1440)$ & 0.64        & 0.35      & 0.01        & 0.00           \\
N3 $S_{11}(1535)$ & 0.50        & 0.08      & 0.42        & 0.00           \\
N4 $S_{11}(1650)$ & 0.56        & 0.20      & 0.16        & 0.08           \\
N5 $D_{13}(1700)$ & 0.12        & 0.75      & 0.10        & 0.03           \\
N6 $P_{11}(1710)$ & 0.10        & 0.50      & 0.30        & 0.10          \\
N7 $P_{13}(1720)$ & 0.11        & 0.81      & 0.03        & 0.05           \\
N8 $D_{15}(1675)$ & 0.45        & 0.53      & 0.01        & 0.01           \\
N9 $F_{15}(1680)$ & 0.65        & 0.35      & 0.00        & 0.00           \\
P1 $P_{11}(1880)$ & 0.06        & 0.55      & 0.37        & 0.02           \\
P2 $P_{13}(1900)$ & 0.08        & 0.73      & 0.08        & 0.11          \\
P3 $F_{15}(2000)$ & 0.08        & 0.88      & 0.04        & 0.00           \\
P4 $D_{13}(1875)$ & 0.08        & 0.90      & 0.01        & 0.01           \\
P5 $F_{15}(1860)$ & --          & --        & --          & --          \\
M1 $D_{13}(2120)$ & 0.10        & 0.90      & 0.00        & 0.00           \\ \hline \hline
\end{tabular}
\caption{Relative branching ratios of nucleon resonances into various channels. These values are used in the computational program of ours and are estimated from Ref.~\cite{PDG16}. For $F_{15}(1860)$, we use the fixed width of 270 MeV from Ref.~\cite{PDG16}.}
\label{tab:BR}
\end{center}
\end{table}

\subsection{Electromagnetic form factors and longitudinal couplings}
In construction of our model for photoproduction~\cite{BS} we kept the explicit dependence on the photon squared mass ($k^2=-Q^2$) in formulae for the scalar and CGLN amplitudes, which allows us to use this formalism also for electroproduction. In the most simple way of extending the model one tends to add only phenomenological form factors in the electromagnetic vertex as it was done, e.g. in the Saclay-Lyon model~\cite{SL}. This naive extension of the model, however, does not work well in the case of the BS1 and BS2 models as they reveal too strong a dependence on the photon mass near the photoproduction point ($Q^2=0$) in the separated $\sigma_T$ and $\sigma_L$ cross sections which is not observed in the data~\cite{Mohring}. Therefore, we added in the formalism also the longitudinal couplings (LC)  for the nucleon resonances to balance this strong $Q^2$ dependence from the transverse couplings (TC). In fitting the free parameters we adopted a rather conservative approach trying to keep the LC coupling constants as small as possible in comparison with the TC ones. Note that the total electroproduction cross section reveals a smooth  dependence for small $Q^2$ in accordance with experimental observations~\cite{Achenbach-sK}.

The electromagnetic structure of hadrons probed by virtual photons with energies considered here can be taken into account via form factors. In this work we include these form factors using the same technique as in Ref.~\cite{SF} which allows us to employ any phenomenological form of the form factors avoiding a violation of the gauge invariance. For the nucleon, hyperons and their resonances, we selected the parametrization by Lomon~\cite{Lomon}, model GKex(02S), which is consistent with vector meson dominance and perturbative QCD in the appropriate momentum transfer regions. For the LC and TC we used the same parametrization. For the $K$, $K^*$, and $K1$, particles exchanged in the $t$-channel we use the expressions of Williams \emph{et al.}~\cite{WJC} and Adelseck-Wright~\cite{AW}, respectively. For the $K^*$ and $K1$ resonances, we assume a monopole form factor,
\begin{equation}
F(Q^2) = \frac{1}{1+Q^2/\Lambda^2},
\label{eq:emff}
\end{equation}
where $\Lambda$ is the cutoff parameter, different for $K^*$ and $K1$, adjusted to experimental data.

The new vertices from the longitudinal coupling of the photon to the proton and  nucleon resonances with spin 1/2, 3/2, and 5/2 considered here are
\begin{equation}
\label{eq:V-EM-1/2}
V^{EM}(N^*_{(1/2)}p\gamma)= -i\frac{g_3^{EM}}{(m_R+m_p)^2}\Gamma_{\mp}\,\gamma_\beta\,{\cal F}^\beta,
\end{equation}
\begin{eqnarray}
\label{eq:V-EM-3/2}
V^{EM}_\mu(N^*_{(3/2)}p\gamma)= -i\frac{g_3^{EM}}{m_R(m_R+m_p)^2}\gamma_5\Gamma_{\mp}\nonumber \\ \times \left(\not\! q\,g_{\mu\beta} -q_\beta\gamma_\mu\right)\,{\cal F}^\beta,
\end{eqnarray}
\begin{eqnarray}
\label{eq:V-EM-5/2}
V^{EM}_{\mu\nu}(N^*_{(5/2)}p\gamma)= -i\frac{g_3^{EM}}{(2m_p)^5}\Gamma_{\mp} (q_\alpha q_\beta g_{\mu\nu} +q^2g_{\alpha\mu}g_{\beta\nu}\nonumber\\
-q_\alpha q_\nu g_{\beta\mu} -q_\beta q_\nu g_{\alpha\mu})\,p^\alpha {\cal F}^\beta,\ \ \ \ \ 
\end{eqnarray}
where $q$ and $p$ are the resonance and proton momenta, respectively, $\Gamma_- = 1$ for negative and  $\Gamma_+ = i\gamma_5$ for positive parity resonances, and ${\cal F}^\beta= k^2 \epsilon^\beta-k\cdot\epsilon\,k^\beta$. 
The latter ensures the gauge invariance and makes the contributions of the Feynman diagrams with the LC proportional to ${Q^2=-k^2}$. 
More details on the formalism for photo- and electroproduction can be found in Ref.~\cite{BS}. Formulae for the new scalar amplitudes are given in the Appendix.  

\section{Discussion of results}
\label{sec:dis}
The free parameters of the model, the coupling constants and cutoff parameters, were adjusted with help of the least-squares fitting procedure using the MINUIT code~\cite{minuit} to the same experimental data for photoproduction used to adjust the BS1 and BS2 model parameters~\cite{BS} supplemented with electroproduction data. The photoproduction data are on differential cross section for $W<2.36\,\text{GeV}$~\cite{AS,Bleck,CLAS05,CLAS10}, hyperon polarization for $W<2.23\,\text{GeV}$~\cite{CLAS10}, and beam asymmetry~\cite{LEPS}. The reason for restricting the photoproduction data sets only to these energies is that in the higher-energy region ($W > 2.4\,\text{GeV}$) more resonances with masses larger than 2 GeV and with the spin higher than 5/2 should be included~\cite{Mart-Sakinah}. The electroproduction data which we have used are the data for unseparated cross section $\sigma_U$~\cite{Brown,Bebek-74,Bebek-77a,Bebek-77b,Azemoon,Mohring}, data for transversal $\sigma_T$ and longitudinal $\sigma_L$ cross sections~\cite{Mohring,Coman,Carman}, data for $\sigma_{LT^\prime}$~\cite{Achenbach-sLTp}, and differential cross section $\sigma_K$~\cite{Achenbach-sK}. Altogether we used 3554 data points in the fitting procedure (3383 and 171 for photoproduction and electroproduction, respectively). For an extensive discussion of the fitting procedure, see Ref.~\cite{BS}.

During the fitting process, the coupling constants of the Born terms, $g_{K\Lambda N}$ and $g_{K\Sigma N}$, were kept inside the limits of broken SU(3) symmetry~\cite{BS} but for the other coupling constants, for the cutoff parameters of hadron form factors, and for the cutoff parameters of electromagnetic form factors of $K^*$ and $K1$ no limitations were imposed. At first, we concentrated on fixing the transverse coupling constants of nucleon resonances fitting to the photoproduction data. Once this was done we extended our database with electroproduction data and, while keeping the transverse coupling constants fixed at the values acquired with help of photoproduction data only, we aimed at finding the optimal values for the longitudinal coupling parameters of $N^*$'s. In the last stage, both transverse and longitudinal couplings of $N^*$'s were released. However, the transverse couplings of $N^*$'s were hardly changed by the handful of electroproduction data, which means that the transverse couplings were very well adjusted solely by the photoproduction data. Let us also note that the coupling parameters of kaon and hyperon resonances were adjusted by the photoproduction data only and then kept constant in the sequel of the fitting procedure.

\begin{table}[h!]
\begin{center}
\begin{tabular}{r r r r r r}
\hline \hline
		                &~value~           &~error~ &                         &~value~           &~error~ \\ \hline
 $g_{K\Lambda N}/\sqrt{4\pi}$   & $-3.00$  & 0.002  &  $G_2(P3)$              &  0.003           & 0.001  \\
 $g_{K\Sigma^0 N}/\sqrt{4\pi}$  &  1.25    & 0.009  &  $G_3(P3)$              & $-0.013$         & 0.005  \\
 $G_V(K^*)$             & $-0.107$         & 0.001  &  $G_1(P4)$              &  0.412           & 0.001  \\
 $G_T(K^*)$             &  0.430           & 0.004  &  $G_2(P4)$              &  0.452           & 0.001  \\
 $G_V(K_1)$             & $-0.177$         & 0.004  &  $G_3(P4)$              &  0.266           & 0.142  \\
 $G_T(K_1)$             & $-0.330$         & 0.007  &  $G_1(P5)$              &  0.014           & 0.001  \\
 $G(N3)$                &  0.227           & 0.004  &  $G_2(P5)$              & $-0.005$         & 0.001  \\
 $G_3(N3)$              & $-0.444$         & 0.468  &  $G_3(P5)$              &  0.067           & 0.007  \\
 $G(N4)$                & $-0.072$         & 0.001  &  $G_1(M1)$              &  0.110           & 0.001  \\
 $G_3(N4)$              & $-1.900$         & 0.235  &  $G_2(M1)$              &  0.087           & 0.001  \\
 $G(N6)$                & $-0.172$         & 0.005  &  $G_3(M1)$              & $-1.995$         & 0.065  \\
 $G_3(N6)$              & $-0.887$         & 0.257  &  $G(L1)$                &  12.790          & 0.040  \\
 $G_1(N7)$              &  0.037           & 0.002  &  $G(L2)$                & $-18.987$        & 0.100  \\
 $G_2(N7)$              &  0.031           & 0.001  &  $G_1(L8)$              & $-1.557$         & 0.050  \\ 
 $G_3(N7)$              &  1.109           & 0.036  &  $G_2(L8)$              &  1.738           & 0.010  \\
 $G_1(N9)$              &  0.017           & 0.002  &  $G_1(S3)$              & $-0.793$         & 0.008  \\
 $G_2(N9)$              & $-0.077$         & 0.002  &  $G_2(S3)$              &  0.126           & 0.002  \\
 $G_3(N9)$              & $-0.578$         & 0.012  &  $\Lambda_{bgr}$        &  1.235           & 0.002  \\
 $G_1(P2)$              & $-0.001$         & 0.002  &  $\Lambda_{res}$        &  0.892           & 0.002  \\
 $G_2(P2)$              & $-0.042$         & 0.001  &  $\Lambda_{K^*}$        &  0.709           & 0.002  \\
 $G_3(P2)$              &  0.195          & 0.060  &  $\Lambda_{K1}$         &  1.503           & 0.026  \\
 $G_1(P3)$              & $-0.003$         & 0.001  &  $\chi^2/\text{n.d.f.}$ &  1.74            & --     \\ \hline \hline

\end{tabular}
\end{center}
\caption{Coupling constants, cutoff values of hadron form factors as well as the electromagnetic ones of $K^*$ and $K1$, and $\chi^2$ of the final model BS3 are displayed. The cutoff values are shown in units of GeV. Errors of the parameters are included as well. For the notation of resonances, we refer to Tab. I. in Ref.~\cite{BS}.}
\label{tab:BS3-par}
\end{table}
We found several outcomes with very similar values of fitted parameters, which probably lie in one deep minimum. According to the smallness of the $\chi^2$, smallness of coupling constants particularly of longitudinal couplings of $N^*$'s and of spin-1/2 hyperon resonances, and the correspondence with data, we selected one of these solutions and coined it BS3 model. Coupling constants, cutoff values, and the $\chi^2$ value of the final BS3 model are summarized in Tab. \ref{tab:BS3-par}. In order to provide information on how sensitive the solution is to variations in the parameters, we include also the errors of fitted parameters. As the longitudinal couplings were fitted only to electroproduction data which are, in quantity and also in quality, inferior to the photoproduction ones, the errors of longitudinal couplings tend to be notably larger than errors of transverse couplings. Concerning the value of the $\chi^2$, it needs to be said that when fitting only to the photoproduction data we were able to get the $\chi^2/\text{n.d.f.}$ value as low as 1.51. It is, therefore, the electroproduction data originating from many different experiments, which worsen the $\chi^2/\text{n.d.f.}$ to its final value of 1.74. This value is, however, still very well acceptable when one considers the $\chi^2/\text{n.d.f.}$ values of different analyses, which tend to be around 1.6 or larger (\emph{e.g.} the multipoles model for $K^+\Lambda$ photoproduction in Ref.~\cite{Mart-Sakinah}).

The numbers of free parameters in the BS models of ours do not differ significantly: there are 31, 28, and 43 parameters adjusted to experimental data in BS1, BS2, and BS3 models, respectively. The increase in the number of free parameters in the BS3 model is mainly due to the inclusion of longitudinal couplings where there is one extra parameter for each nucleon resonance (and since we include 10 $N^*$'s in the BS3 model there are 10 longitudinal couplings of theirs to be adjusted). With help of these longitudinal couplings of virtual photons to nucleon resonances we are able to achieve a reasonable description of electroproduction data. This, however, could not be done without these couplings; see \emph{e.g.} the prediction of BS1 model, where there are no LC implemented, which falls short in reproducing experimental data for $Q^2\neq 0$ (how the BS1 model works at very small $Q^2$ can be also seen in Ref.~\cite{Achenbach-sLTp} where its predictions of $\sigma_{LT^\prime}$ are shown and compared to data). In the beginning of our fitting procedure, we kept the cutoff parameters of electromagnetic form factors of $K^*$ and $K1$ fixed at values found in the Kaon-MAID analysis, \emph{i.e.} $\Lambda_{K^*}=1.51\,\text{GeV}/c$ and $\Lambda_{K1}=0.67\,\text{GeV}/c$. However, we realized that with these values of cutoff parameters we are not able to obtain a dependable description of some electroproduction data in certain kinematic regions. We, therefore, released these parameters from their Kaon-MAID values and found different values for both cutoff parameters which allow an acceptable description of electroproduction data. 

The content of $N^*$ states in the BS3 model overlaps well with other BS models. There are 8, 9, and 10 nucleon resonances in BS1, BS2, and BS3 models, respectively. The nucleon resonances included in the BS1 model can be found also in BS2 and BS3 models. On top of that, the extra nucleon resonance in the BS2 model is the $P_{11}(1710)$ state and the two extra $N^*$ states in the BS3 models are $P_{11}(1710)$ and $D_{13}(2120)$ states. Only a few coupling parameters of nucleon resonances remain on roughly the same values in both BS1 and BS3 models. The most notable changes occur in the coupling parameters of $S_{11}(1535)$ and $D_{13}(1875)$ which are twice as large in the BS3 model as in the BS1 model and couplings of $P_{13}(1720)$ and $F_{15}(1860)$ which change their signs. Changes can be found also in the couplings of kaon resonances: while the tensor coupling of $K^*$ is an order of magnitude larger in the BS3 model in comparison with BS1 model, the vector coupling of $K1$ is in the same comparison twice smaller with opposite sign. This consequently means that various resonances interfere in a different way, which then changes also the dynamics of the model.

Whereas the set of $N^*$'s is more or less settled, the selection of $Y^*$'s (not only) in our models differs substantially. Interestingly, there is no hyperon resonance which can be found in all of the BS models and there is only one $Y^*$, namely the $\Lambda(1890)$, that can be found in both BS1 and BS3 models (for the resonances included in BS1 and BS2 models, see Tab. II. in Ref.~\cite{BS}). This means that the most significant $Y^*$'s still can not be unambiguously determined with help of currently available data.

\begin{figure}
\includegraphics[angle=270,width=\columnwidth]{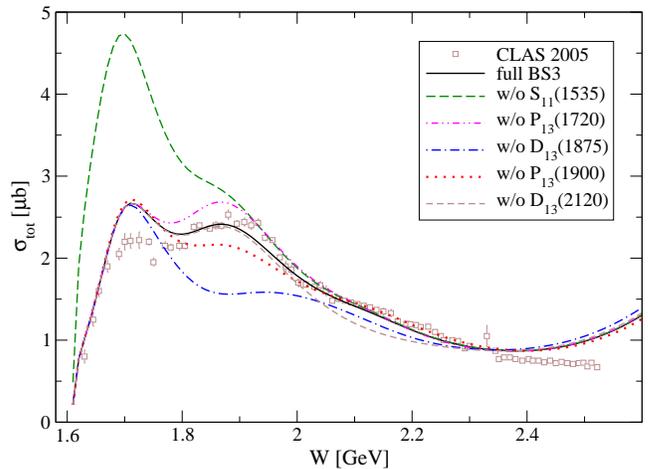}
\caption{Total cross section as predicted by the BS3 model (solid curve) in comparison with CLAS 2005 data~\cite{CLAS05}. The other curves illustrate the behaviour of the BS3 model without particular resonance.}
\label{fig:totcrs-N}
\end{figure}

\begin{table}[h!]
\begin{center}
\begin{tabular}{l r l r}
\hline \hline
   & $\Delta \chi^2$ [\%] & & $\Delta \chi^2$ [\%]  \\ \hline
$S_{11}(1535)$-N3 & 331  & $P_{13}(1900)$-P2 &  826  \\
$S_{11}(1650)$-N4 &  81  & $F_{15}(2000)$-P3 &   30  \\
$P_{11}(1710)$-N6 &  43  & $D_{13}(1875)$-P4 &  844  \\
$P_{13}(1720)$-N7 & 188  & $F_{15}(1860)$-P5 &   82  \\
$F_{15}(1680)$-N9 & 202  & $D_{13}(2120)$-M1 &  125  \\ \hline \hline
\end{tabular}
\end{center}
\caption{Effect of the nucleon resonances on reducing the $\chi^2$. The values shown are defined in the text.}
\label{tab:N-chi2}
\end{table}
After finishing the fitting procedure, we wanted to estimate the role played by a particular nucleon resonance in reducing the $\chi^2$ value, which determines its importance. We, therefore, computed the $\chi^2$ omitting the $N^*$ states one after another. The values shown in Tab. \ref{tab:N-chi2} are calculated as
\begin{equation}
\Delta\chi^2 = \frac{\chi^2_{N^*} - \chi^2}{\chi^2}\cdot 100\%,
\label{eq:Dchi2}
\end{equation}
where $\chi^2_{N^*}$ and $\chi^2$ denote the value of $\chi^2$ with the particular resonance omitted and the value of the best fit, respectively. Note that in this discussion and in Tab. \ref{tab:N-chi2}, $\chi^2$ is a shorthand notation for $\chi^2/\text{n.d.f.}$. We need to mention that a relation similar to Eq.~(\ref{eq:Dchi2}) was used also in Ref.~\cite{Mart-Sakinah} and the purpose of this relation is to give us information on how difficult it would be to aptly depict the experimental data without particular nucleon resonance. In order to illustrate the values shown in Tab.~\ref{tab:N-chi2}, we include a prediction of the total cross section by the BS3 model, see Fig.~\ref{fig:totcrs-N}, where some of the nucleon resonances affecting the $\chi^2$ most strongly are omitted. According to the Tab.~\ref{tab:N-chi2}, the most important nucleon resonances to get a reliable description of experimental data are the $P_{13}(1900)$ and $D_{13}(1875)$ states which shape the model predictions especially around the second peak in the cross section, \emph{i.e.} around $W=1.9\,\text{GeV}$ (the former was also found to be a basic resonance in the multipole analysis~\cite{Mart-Sakinah}). The total-cross-section prediction corroborates this since it reveals a constructive interference of these resonances with other terms in the energy region $W = 1.8-2.0\,\text{GeV}$. What is more, the Fig.~\ref{fig:totcrs-N} shows an important destructive interference of $P_{13}(1720)$ state in the same energy region even though its omission does not lead to as significant an increase of the $\chi^2$ as the omission of $P_{13}(1900)$ and $D_{13}(1875)$ states. Worth mentioning is also the $S_{11}(1535)$ state which, being a dominant nucleon resonance in the case of $\eta$ photoproduction, produces on its own the shape of the $K^+\Lambda$ photoproduction cross section right at the threshold. The coupling strength of this resonance in the $K^+\Lambda$ channel was also investigated in Ref.~\cite{Mart2013} using an isobar model. It is interesting that coupling constants extracted in that analysis are in the range 0.16--0.29 which agrees very well with our value 0.23. When this resonance is left out from our model, we get a prediction of total cross section largely overshooting the experimental data and also the full model prediction. This striking result can be explained easily since the total cross section shows an integral effect and the $S_{11}(1535)$ state interferes destructively with other terms at all kaon angles. Its effects at various angles, therefore, add coherently and one can then observe such a dramatic increase in the total cross section in the threshold region once this state is omitted.
Notable is also the presence of $D_{13}(2120)$ nucleon resonant state since it is one of the key resonances to match the data according to Ref.~\cite{Schumacher-Sargsian} pointing to the structure near $W=2.1\,\text{GeV}$ (visible in the differential cross section mainly in the forward-angle region and also in the total cross section, Fig.~\ref{fig:totcrs-N}). This state influences the reaction below 2 GeV as well.

%
%

In the BS3 model, the hadron form factor of dipole shape is implemented since use of other shapes of hadron form factor lead to significantly larger values of cutoff parameters, $\chi^2/\text{n.d.f.}$ and fitted coupling constants. In our analysis, we revealed that even such a weak form factor as the dipole one is can be used to sufficiently suppress the contributions of resonances with various spin. However, a necessary prerequisite for this statement are small cutoff parameters (in the BS3 models they acquire values of $1.24\,\text{GeV}$ and $0.89\,\text{GeV}$ for background and resonant terms, respectively). These cutoff values are much smaller than in the BS2 model where the dipole shape of the hadron form factor is exploited as well. We deem this to be an effect of energy-dependent widths: the resonance width increases with energy, which means the resonance is farther from the physical region and beyond the resonance region consequently contributes less. However, the widths in the resonance region may be very small in comparison with the values inserted while assuming fixed decay widths, which leads to very strong contributions that need to be suppressed by the use of strong hadron form factor.

The electromagnetic form factors of $K^*$ and $K1$ tend to fall rapidly with $Q^2$ as is indicated by the inclination of their cutoff parameters to, in many of our fits, unrealistically small values. Moreover, we noticed that stronger electromagnetic form factors for $K^*$ and $K1$ also lead to a reduced $\chi^2$ value and are, thus, preferred by the data. A remedy would be to opt for a stronger shape of the form factor, which could then lead to a higher value of cutoff parameters. Instead of introducing a different form factor for $K^*$ and $K1$ resonances, we turned the condition on realistic cutoff parameters of $K^*$ and $K1$ electromagnetic form factors into one of the most stringent criteria on choosing particular result as our new model. The resulting cutoff parameters are $\Lambda_{K^*}=0.71\,\text{GeV}$ and $\Lambda_{K1}=1.50\,\text{GeV}$ for $K^*$ and $K1$, respectively.

\subsection{Photoproduction}
\begin{figure} 
  \includegraphics[angle=270,width=\columnwidth]{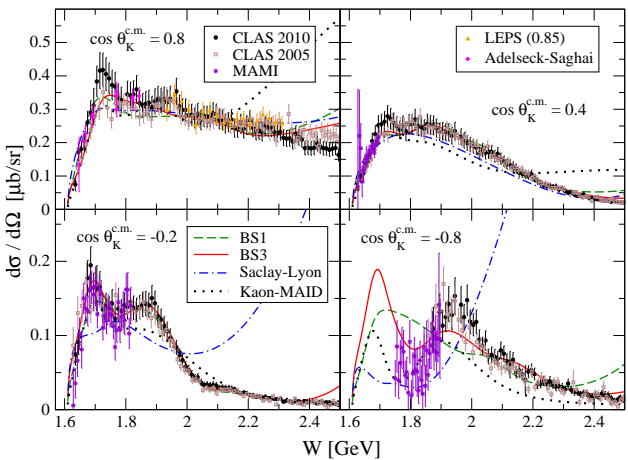}
\caption{Cross-section predictions of BS1 (dashed curve), BS3 (solid curve), Saclay-Lyon (dash-dotted curve), and Kaon-MAID (dotted curve) models are shown for four kaon center-of-mass angles. The data are from CLAS 2005~\cite{CLAS05}, CLAS 2010~\cite{CLAS10}, MAMI~\cite{MAMI-C}, and LEPS~\cite{LEPS} collaborations and from Ref. \cite{AS}.}
\label{fig:crs-w}
\end{figure}

\begin{figure} 
\includegraphics[angle=270,width=\columnwidth]{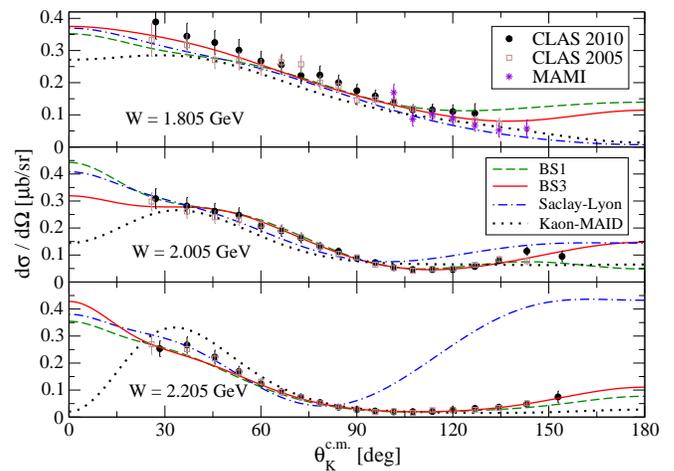}
\caption{Angular dependence of the cross section is shown for three values of the c.m. energy. The data are from CLAS 2005~\cite{CLAS05}, CLAS 2010~\cite{CLAS10}, and MAMI~\cite{MAMI-C} collaborations. Notation of the curves is the same as in the Fig.~\ref{fig:crs-w}.}
\label{fig:crs-th0}
\end{figure}

\begin{figure}
\includegraphics[angle=270,width=\columnwidth]{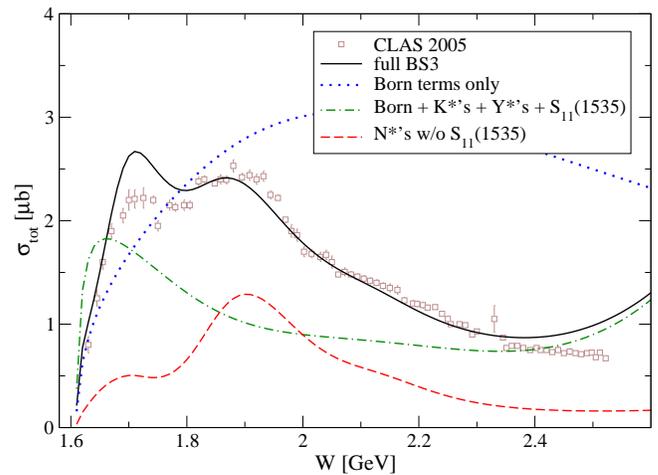}
\caption{Total-cross-section prediction of the BS3 model (solid curve) compared to CLAS 2005 data~\cite{CLAS05}. Predictions of sole Born terms (dotted curve), background terms (dash-dotted curve), and nucleon resonances (dashed curve) are shown.}
\label{fig:totcrs-sut}
\end{figure}

\begin{figure} 
\includegraphics[angle=270,width=\columnwidth]{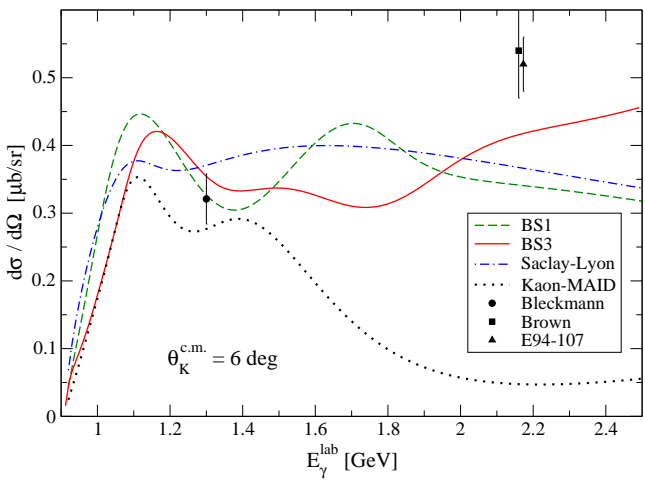}
\caption{Results for the differential cross section for $p(\gamma,K^+)\Lambda$ at $\theta_K^{c.m.}=6^\circ$ are shown for several models. Data points of Brown~\cite{Brown} and E94-107~\cite{E94-107} are for electroproduction with a very small value of the virtual-photon mass $|k^2|$; the only photoproduction datum available in this region stems from Bleckmann \emph{et al.}~\cite{Bleck}. Notation of the curves is the same as in the Fig.~\ref{fig:crs-w}.}
\label{fig:crs-6deg}
\end{figure}

\begin{figure} 
\includegraphics[angle=270,width=\columnwidth]{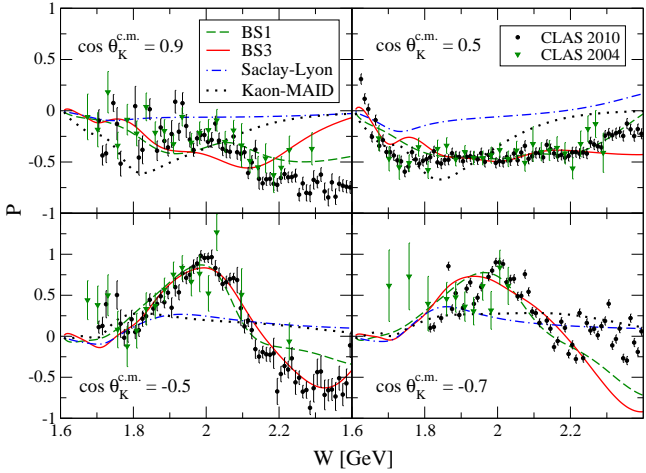}
\caption{Results for the hyperon polarization $P$ are shown for several kaon c.m. angles. Data stem from CLAS collaboration, Refs.~\cite{CLAS10} and ~\cite{CLAS04}. Notation of the curves is the same as in the Fig.~\ref{fig:crs-w}.}
\label{fig:pol-w}
\end{figure}

\begin{figure} 
  \includegraphics[angle=270,width=\columnwidth]{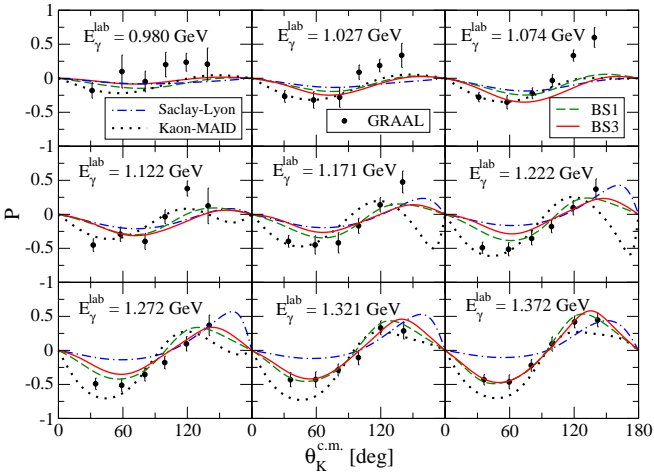}
\caption{Predictions of hyperon polarization $P$ are shown for several values of energy $E_\gamma^{lab}$. Data stem from GRAAL collaboration, Ref.~\cite{GRAAL-P}. Notation of the curves is the same as in the Fig.~\ref{fig:crs-w}.}
\label{fig:pol-th0-GRAAL}
\end{figure}

\begin{figure} 
\includegraphics[angle=270,width=\columnwidth]{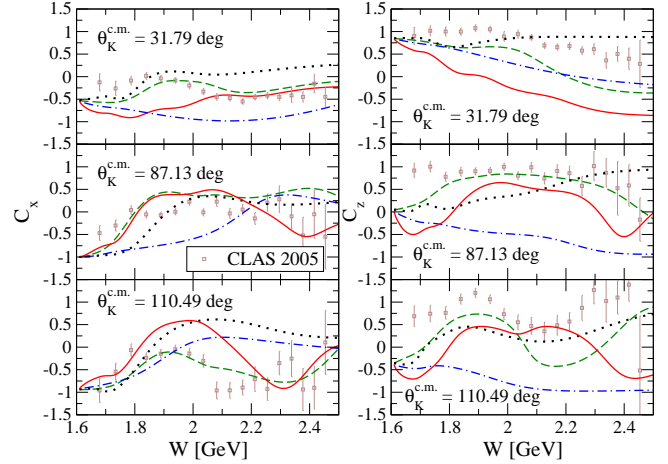}
\caption{Predictions for the double-polarization observables $C_x$ and $C_z$ are shown for various kaon c.m. angles. Notation of the curves is the same as in the Fig.~\ref{fig:crs-w} and the data stem from the CLAS analysis~\cite{CLAS-double}.}
\label{fig:cxcz}
\end{figure}

\begin{figure} 
\includegraphics[angle=270,width=\columnwidth]{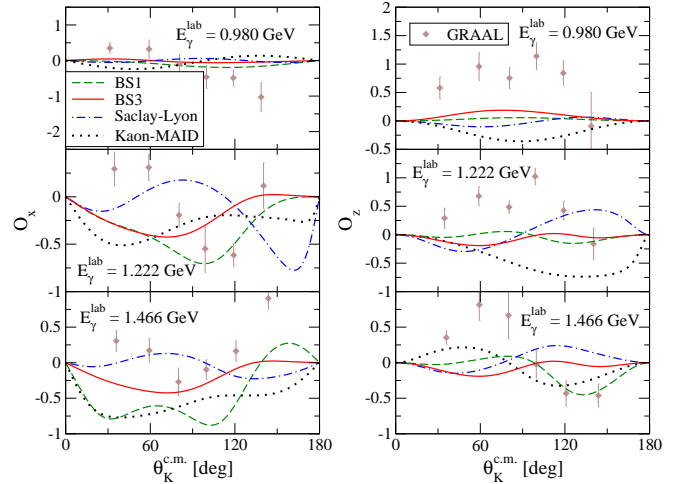}
\caption{Double-polarization observables $O_x$ and $O_z$ are shown for various kaon c.m. angles. Notation of the curves is the same as in the Fig.~\ref{fig:crs-w}. The data are from the GRAAL collaboration~\cite{GRAAL}. The errors are quadratic sums of statistical and systematic uncertainties.}
\label{fig:oxoz}
\end{figure}

The results of the BS3 model for photoproduction are shown in Figs.~\ref{fig:crs-w}--\ref{fig:oxoz} in comparison with data and predictions from other models. All model predictions presented in this paper have been obtained with our computational code which gives us outcomes of various models which are perfectly mutually consistent.
The predicted cross section at $\cos \theta_{K}^{c.m.}=0.8$, Fig.~\ref{fig:crs-w}, reveals a two-peak structure at forward angles. The first peak is shaped mainly by $S_{11}(1535)$ and $S_{11}(1650)$ contributions whereas the apparent structure around $W = 2\,\text{GeV}$ at angles in the forward hemisphere is caused mainly by $D_{13}(2120)$ and $F_{15}(1685)$ and their interference with other terms. Contribution of $P_{11}(1710)$ is the most important one for the creation of the peak at $W=1.75\,\text{GeV}$, while the interference of $P_{13}(1900)$ and $D_{13}(1875)$ with other terms shapes the cross section primarily in the energy range $W=1.8-2.0\,\text{GeV}$. We also point out the apparent inconsistency in data from CLAS 2010~\cite{CLAS10} and CLAS 2005~\cite{CLAS05} near the threshold where the former data produce much sharper peak than the latter ones. All of the models shown in Fig.~\ref{fig:crs-w} prefer the CLAS 2005 data. As was the case in the BS1 and BS2 model, the $S_{11}(1650)$ is not strong enough to produce a narrow structure near the threshold even in the BS3 model and its predictions are, therefore, in concert rather with the older CLAS data.
%
%

In the cross-section prediction of BS3 model at $\cos \theta_{K}^{c.m.}=0.4$, there is an interesting peak at $W=1.85\,\text{GeV}$, which is most probably caused by constructive interference of $D_{13}(1875)$ with other terms. This high-spin resonance thus ``resonates" at its mass, showing that, at least for the spin-3/2 resonance, the dipole shape of the hadron form factor with $\Lambda_{res}\approx 1\,\text{GeV}$ is strong enough to cut off its contribution far from the resonance energy. Let us note that the situation depicted in the Fig. 2 in the Ref.~\cite{BS} is an exaggeration: in fact, the coupling constants of $N^*(5/2)$ states usually acquire one or even two magnitudes lower values than the ones shown in the mentioned figure and the hadron form factors with given cutoff values are, therefore, much more effective in taming their contributions. The other most important contribution in this energy region comes from $P_{13}(1900)$.
%
%

%
%

The predicted angular dependence of the cross section is compared to the experimental data in the Fig.~\ref{fig:crs-th0} for three energies. The model predictions differ mainly in the forward-angle region, where the Kaon-MAID model predicts suppressed cross section at larger energies, and in the backward-angle regions, where the Saclay-Lyon overshoots the data at $W=2.205\,\text{GeV}$. The prediction of the cross-section at $W=2.005\,\text{GeV}$ by the BS3 model is notably smaller than the prediction by the BS1 model, as in the BS3 model the spin-5/2 nucleon resonances suppress the contribution of background terms at the forward angles. The spin-3/2 nucleon resonances contribute mainly in the central-angle region but when combined with the background terms they contribute strongly also in the forward-angle region. Moreover, spin-1/2 and spin-3/2 $N^*$'s suppress background contributions at backward angles and energies below 2 GeV. All of this illustrates how important the interference terms can be for the correct depiction of cross sections.

In Figure~\ref{fig:totcrs-sut} we show various contributions to the 
total cross section from the non-resonant and resonant parts of the 
amplitude. It is well known that the Born contributions provide too big 
cross sections, especially for energies above 2 GeV. 
These contributions are significantly reduced by the resonance 
contributions in the $t$- and $u$-channels (see the dashed line in 
Fig.~\ref{fig:totcrs-sut}). The $s$-channel 
resonance $S_{11}(1535)$ is included 
into the  background part instead of among the $N$*'s as its pole is 
located below the K$\Lambda$ threshold. However, since this pole is near the threshold its contribution 
interferes destructively with other background terms reducing 
the peak at the threshold as shown in Fig.~4  
(see also Fig.~\ref{fig:totcrs-N} and the corresponding discussion on $S_{11}(1535)$ ). 
The $s$-channel resonances are important mainly in the $1.8-2.1$ GeV energy region.

In the very forward-angle region, where the data is scarce, differences among  predictions of different models are immense, particularly for $E_\gamma^{lab} > 1.5\,\text{GeV}$, see Fig.~\ref{fig:crs-6deg}. The BS models together with Saclay-Lyon model predict similar magnitude of the cross section, while the Kaon-MAID model is for $E_\gamma^{lab} > 1.5\,\text{GeV}$ suppressed owing to suppression of the proton exchange by strong hadron form factors. In the BS3 model, the $D_{13}(1875)$ state seems to be the most prominent resonant contribution in creating the plateau-like behaviour in the energy range $E_\gamma^{lab}=1.4-1.7\,\text{GeV}$. It is the $D_{13}(2120)$ state destructive interference with other terms, which shapes the cross-section prediction at around $E_\gamma^{lab}=1.8\,\text{GeV}$. Moreover, note that the BS3 model predicts larger cross section than the BS1 version at energies above 2 GeV. Its prediction is, therefore, closer to the recent data point from the JLab experiment E94-107~\cite{E94-107}.
%
%

In the Figs.~\ref{fig:pol-w}--\ref{fig:oxoz}, predictions for spin observables are given and they are compared to CLAS and GRAAL data. For energy dependence of hyperon polarization $P$, the BS models are in concord with experimental data in the kinematic region where they were fitted to data, \emph{i.e.} from the threshold up to approximately 2.2~GeV. On the other hand, Kaon-MAID and Saclay-Lyon models do not capture the shape of data as they were not fitted to these data. Fig.~\ref{fig:pol-th0-GRAAL} shows comparison of model predictions for hyperon polarization with data from GRAAL collaboration. Since none of the models shown was fitted to GRAAL polarization data, this figure shows merely the predictive power of the models. In the central-angle region, all models capture the shape of data for all energies quite well. Some discrepancies can be seen only at the forward and backward kaon angles but altogether the models depict the data aptly.

Similarly to GRAAL hyperon polarization data, none of the models shown in this work was fitted to experimental data on double-polarization observables $C_x$, $C_z$, $O_x$, and $O_z$. These figures, thus, again collect mere predictions of the models. The Saclay-Lyon model does not work well for $C_z$, especially for large kaon angles, where its prediction is negative while the data are above zero. Other models give better results as their predictions are of the same sign as the data and the BS1 model fits also the shape of the data. However, predictions of $C_z$ by the BS3 model are in concert with data only in some kinematic regions, whereas for large kaon angles and near threshold and above approximately $2.3\,\text{GeV}$ they have opposite sign than data. For $O_x$ and $O_z$ there are much less data available when compared to $C_x$ and $C_z$. Nevertheless, the models are not able to capture data satisfactorily, some of the models even produce predictions with an opposite sign in comparison with data (such as the SL and BS3 model prediction of $O_z$ for large $E_\gamma^{lab}$). Only in a few kinematic regions there can be found a cursory correspondence between some of the models and data.

%
%

%
%

%
%

%
%

\begin{figure} 
\includegraphics[width=\columnwidth]{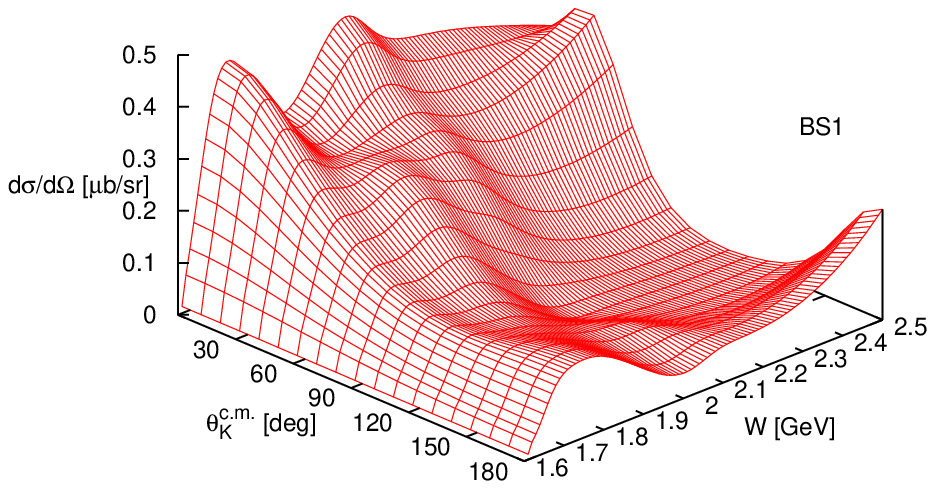}
\caption{Overall description of the resonance region of the $p(\gamma,K^+)\Lambda$ process by the BS1 model for all kaon c.m. angles and for energy from the threshold up to 2.5~GeV.}
\label{fig:BS1-3D}
\end{figure}

\begin{figure} 
\includegraphics[width=\columnwidth]{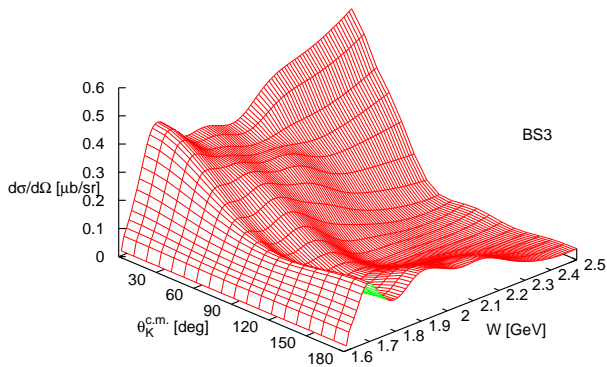}
\caption{The same as Fig.~\ref{fig:BS1-3D} but with BS3 model.}
\label{fig:BS3-3D}
\end{figure}

In Figs.~\ref{fig:BS1-3D} and~\ref{fig:BS3-3D}, we summarize the behaviour of our new isobar models in the resonance region, \emph{i.e.} for $W$ from the threshold up to 2.5~GeV and for all kaon c.m. angles $\theta_K^{c.m.}$. In the BS1 model (Fig.~\ref{fig:BS1-3D}), two peaks develop right at the zero kaon angle and prevail the cross-section depiction at forward kaon angles. A third peak emerges between them from $\theta_K^{c.m.}\approx 40^\circ$ and these three peaks then constitute the cross-section prediction for central angles. In the backward angles, \emph{i.e.} for angles larger than $120^\circ$, the cross section is dominated by only one very broad peak. The cross section description provided by the BS3 model is slightly different, see Fig.~\ref{fig:BS3-3D}. There is only one peak for angles $\theta_K^{c.m.}=0^\circ-30^\circ$ while the peak seen in the BS1 model prediction at $W=2\,\text{GeV}$ is missing and a plateau from 1.8~GeV to 2.1~GeV is created instead. In the central kaon angles, three peaks can be recognized, similarly to the BS1 model, which are visible also at backward angles, even though two of them are very broad.

%
%

%
%

\subsection{Electroproduction}
\begin{figure} 
  \includegraphics[angle=270,width=\columnwidth]{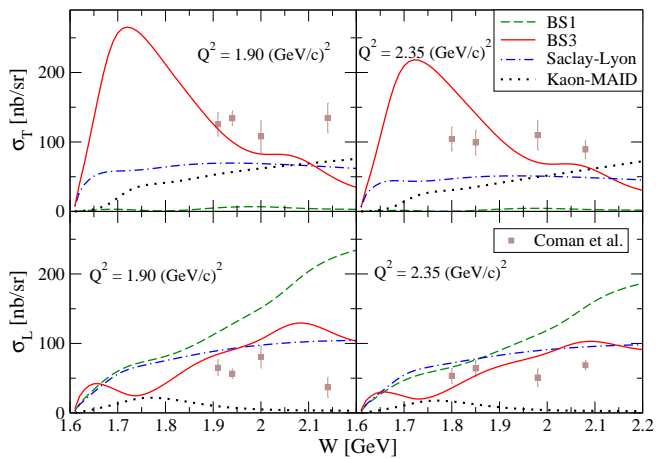}
\caption{Energy dependence of transverse, $\sigma_T$, and longitudinal, $\sigma_L$, cross sections for $Q^2=1.90\,(\text{GeV}/c)^2$ and $Q^2=2.35\,(\text{GeV}/c)^2$ and for zero kaon angle is shown. The result of the BS3 model and predictions of BS1, Saclay-Lyon, and Kaon-MAID models are compared with JLab  data~\cite{Coman}. Notation of the curves is the same as in the Fig.~\ref{fig:crs-w}.}
\label{fig:sLT-c}
\end{figure}

\begin{figure} 
\includegraphics[angle=270,width=\columnwidth]{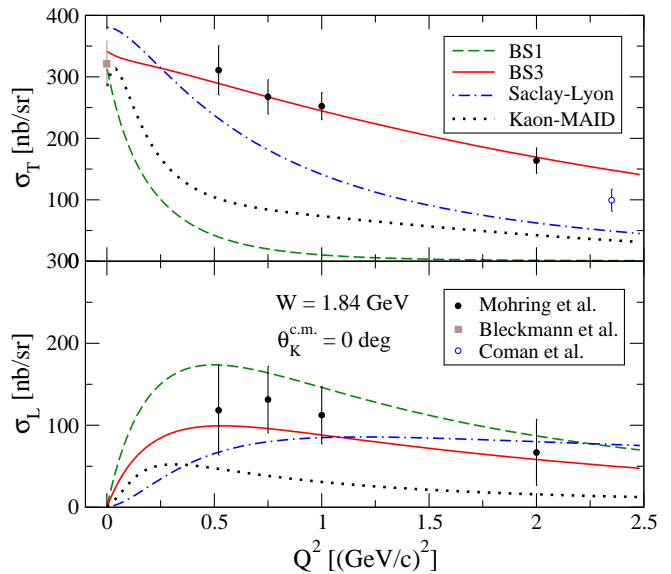}
\caption{Transverse, $\sigma_T$, and longitudinal, $\sigma_L$, cross sections for kaon electroproduction at $W=1.84\,\text{GeV}$ and for zero kaon angle as function of $Q^2$ are shown. The result of the BS3 model and predictions of BS1, Saclay-Lyon, and Kaon-MAID models are compared with JLab~\cite{Mohring,Coman} and Bleckmann {\it et al.}~\cite{Bleck} ($Q^2=0$ for $\sigma_T$) data. Notation of the curves is the same as in the Fig.~\ref{fig:crs-w}.}
\label{fig:sLT-m}
\end{figure}

\begin{figure} 
\includegraphics[angle=270,width=\columnwidth]{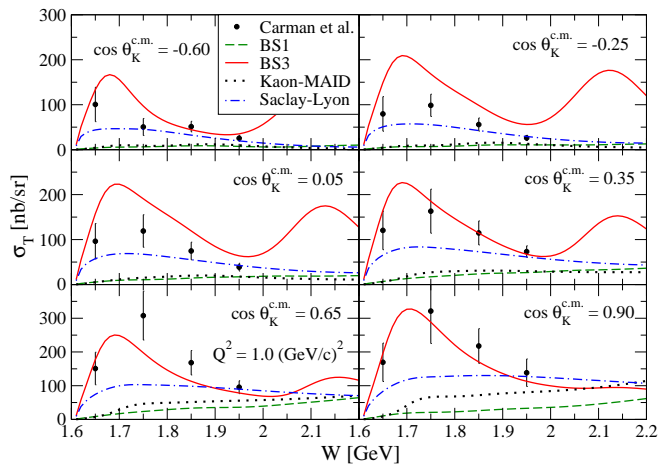}
\caption{Energy dependence of transverse cross section $\sigma_T$ at $Q^2 = 1.0\,(\text{GeV}/c)^2$ is shown for several kaon angles. We compare the result of the BS3 model and predictions of BS1, Saclay-Lyon, and Kaon-MAID models with CLAS data~\cite{Carman}. Notation of the curves is the same as in the Fig.~\ref{fig:crs-w}.}
\label{fig:sT}
\end{figure}

\begin{figure} 
\includegraphics[angle=270,width=\columnwidth]{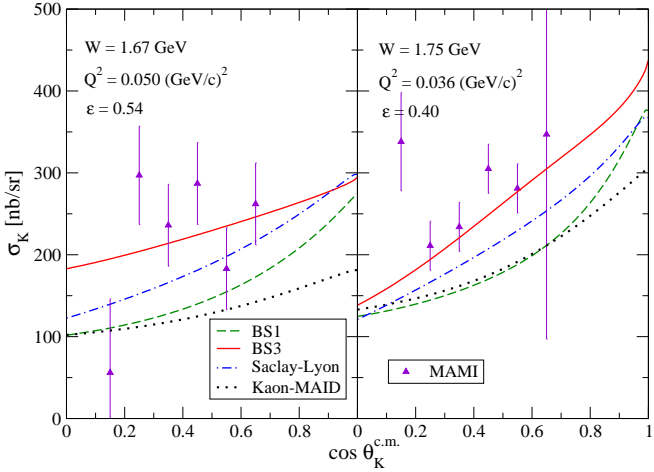}
\caption{Differential cross section of kaon electroproduction in two kinematic regions is shown. The result of the BS3 model and predictions of BS1, Saclay-Lyon, and Kaon-MAID models are compared with MAMI data~\cite{Achenbach-sK}. Notation of the curves is the same as in the Fig.~\ref{fig:crs-w}.}
\label{fig:sK}
\end{figure}

In Figs.~\ref{fig:sLT-c}--\ref{fig:sK}, we provide results of the BS1, BS3, Saclay-Lyon, and Kaon-MAID models for electroproduction and compare them with available data. Note that the data for $K^+\Lambda$ electroproduction are not comparable to those for $K^+\Lambda$ photoproduction, either in number or in quality. Except for the BS3 model, no model was fitted to the data shown. Let us also note that in the calculation with the Kaon-MAID model we do not assume longitudinal couplings and it may be thus reasonable to call the model reduced Kaon-MAID model where calculations for nonzero $Q^2$ are concerned. 

The energy dependence of separated cross sections is given in Fig.~\ref{fig:sLT-c} for two values of $Q^2$. Whereas the BS3 model captures the magnitude and shape of both $\sigma_T$ and $\sigma_L$ data, and Saclay-Lyon predictions are of the same magnitude as $\sigma_T$ data, reduced Kaon-MAID and BS1 models fail to reproduce the data on both $\sigma_T$ and $\sigma_L$. The Saclay-Lyon model produces structureless dependence of $\sigma_T$ and $\sigma_L$ on $W$ while the BS3 model gives a resonance-like shape of both $\sigma_T$ and $\sigma_L$. The peak near the threshold of $\sigma_T$ is created mainly by LC contributions of $N^*(1/2)$ states, although the most important contribution stems from $N^*(3/2)$ states whose LC contributions predominate the $\sigma_T$ prediction by the BS3 model for energies above 1.9~GeV. The spin-5/2 nucleon resonances, on the other hand, slightly suppress the $\sigma_T$ at high energies by their destructive interference with other terms. The effect of longitudinal contributions for description of $\sigma_L$ is not as striking as for description of $\sigma_T$. Nevertheless, they are important as they help capture the magnitude and, in part, also the shape of $\sigma_L$ data. The most significant is the LC contribution of $N^*(1/2)$'s, which creates the peak near the threshold and also suppresses the $\sigma_L$ prediction in the whole energy range shown. Note that in this kinematic region with large $Q^2$, results might be also sensitive to a choice of the electromagnetic form factors. 
 
In the Fig.~\ref{fig:sLT-m}, the dependence of transverse, $\sigma_T$, and longitudinal, $\sigma_L$, cross sections on $Q^2$ for $W=1.84\,\text{GeV}$ and for zero kaon angle is shown. For a correct description of transverse cross section $\sigma_T$, the inclusion of longitudinal couplings for $N^*$'s is absolutely vital in our approach. What happens when one omits these longitudinal couplings is aptly illustrated by the BS1 model where there are no longitudinal couplings included and the cross section falls steeply with $Q^2$. The dependence of $\sigma_T$ on $Q^2$ is, thus, almost entirely given by the longitudinal couplings of nucleon resonances to the virtual photon. This explains the noteworthy behaviour of the BS3 model, which falls steadily  with $Q^2$ even more slowly 
than the Saclay-Lyon model. The contributions of longitudinal couplings of $N^*$ states are proportional to $Q^2$ (see formulae for the scalar amplitudes in the Appendix where $Q^2=-k^2$) and they are, thus, negligible at very small $Q^2$ where contributions of transverse couplings dominate. The prediction of $\sigma_T$ by the BS3 model, therefore, reveals a harmonious interplay between transverse and longitudinal couplings. On the other hand, as the BS1-model prediction in Fig.~\ref{fig:sLT-m} illustrates, we are able to get a reliable description of data on $\sigma_L$ even with no LC contributions in some kinematic regions.

Interestingly, all LC contributions of $N^*$'s are of almost the same importance for a reasonable description of $\sigma_T$ cross section at $W=1.84\,\text{GeV}$. For the capture of $\sigma_T$ data, the omission of LC contributions of either $N^*(1/2)$'s, $N^*(3/2)$'s or $N^*(5/2)$'s leads to a suppressed $\sigma_T$ prediction which is nearly at the level of model prediction with no LC contributions at all. On the other hand, leaving out LC couplings of $N^*$ states with various spin has various effects on $\sigma_L$ description.  The LC contributions of $N^*(1/2)$'s, on the one hand, help suppress the excessively large values of $\sigma_L$ given solely by the transverse couplings of $N^*$'s. The role of the LC coupling of $N^*(5/2)$ states, on the other hand, lies in supporting the $\sigma_L$ prediction by interfering constructively with other terms as the omission of $N^*(5/2)$ LC couplings leads to $\sigma_L$ prediction suppressed by approximately 50~nb/sr. Due to nearly balanced effects from the $N^*(1/2)$ and $N^*(5/2)$ states the LC contributions of $N^*(3/2)$ states therefore make a net effect of the longitudinal couplings in $\sigma_L$.

In the Fig.~\ref{fig:sT}, we present results of the transverse cross section $\sigma_T$ for several kaon center-of-mass angles $\theta_K^{c.m.}$. As only the BS3 model was fitted to the CLAS data shown in the Fig.~\ref{fig:sT}, it is the only model capable of capturing their shape, particularly at forward angles. Saclay-Lyon model can produce $\sigma_T$ prediction with a shape similar to experimental data for backward and central kaon angles, while it fails to reproduce the peak at forward angles. With no longitudinal couplings, the BS3 model produces an outcome similar to the BS1 model, \emph{i.e.} a smooth cross section with no apparent structure. The two significant peaks are thus created solely by the longitudinal couplings of nucleon resonances to virtual photons. The first peak is produced mainly by the LC contribution of $N^*(1/2)$ states where the $S_{11}(1650)$ state plays the primary role, while the largest contribution to the second peak stems from longitudinal couplings of $N^*(3/2)$'s, particularly of the $D_{13}(2120)$ state. Note that the second peak does not appear at small kaon angles. This resonance pattern is similar to what we observe in the upper part of Fig.~\ref{fig:sLT-c} as there is also shown the energy dependence of $\sigma_T$ but for a different $Q^2$ value and zero kaon angle.

In the Fig.~\ref{fig:sK}, the results for differential cross section $\sigma_K$ are given for two kinematic regions (for the definition of $\sigma_K$ we refer to Eq.~(22) in Ref.~\cite{BS}). None of the models gives utterly unacceptable predictions although BS1 and reduced Kaon-MAID models predict smaller $\sigma_K$ than other models which lie below the experimental data. The Saclay-Lyon model works slightly better as its calculated $\sigma_K$ is closer to data. The description provided by the BS3 model is in accordance with experimental data, particularly at $W=1.75\,\text{GeV}$ where the data are less scattered and a pattern for steeply rising $\sigma_K$ with cosine of the kaon angle can therefore be seen. Longitudinal couplings of nucleon resonances play an important role for a sound description of all electroproduction response functions and $\sigma_K$ is no exception as without them we would get a suppressed $\sigma_K$ prediction (similar to the one given by the BS1 model). The most important LC contribution comes from $N^*(1/2)$ states since when these contributions are omitted the $\sigma_K$ drops by approximately 100~nb/sr and it is even smaller than the $\sigma_K$ produced solely by transverse couplings. On the other hand, the LC contributions of spin-3/2 and spin-5/2 nucleon resonances have only a slight influence on the $\sigma_K$ prediction, which is tangible mainly at forward angles where they interfere constructively and destructively, respectively, with other terms. 
Note that in this case with very small value of $Q^2$ [$Q^2=0.05\ (\text{GeV}/c)^2$] 
the analysis of LC almost does not depend on a choice of the electromagnetic 
form factors contrary to the discussion of results in Figs.~\ref{fig:sLT-c} and \ref{fig:sT} with $Q^2$ larger than 1~$(\text{GeV}/c)^2$. 

%
%

%
%

%
%

%
%

\section{Conclusion}
\label{sec:con}
We have constructed a new version of our isobar model, BS3, 
that reveals a considerable improvement as to description of electroproduction data. Beside the longitudinal couplings and electromagnetic form factors, in the BS3 we have also included 
energy-dependent widths of the nucleon resonances that can be considered to be a partial restoration of unitarity in our single-channel approach. A significant improvement with respect to the BS1 model, extended in the simple fashion by considering only the electromagnetic 
form factors, is observed for the separated cross sections at small kaon angles and small virtual-photon mass ($Q^2$), which is displayed in comparison with JLab and MAMI data. 
 
In electroproduction, the main difference between the BS3 and extended BS1 models lies in the inclusion of the LC in the BS3 model. The longitudinal couplings appear to be highly important for a reasonable description of the data, especially at small values of $Q^2$ where the electromagnetic form factors still do not play a significant role. Values of the LC, $G_3$, acquired moderate values with respect to the transversal coupling constants, $G_1$ and $G_2$, which we regard as a reasonable and realistic extension of the model for nonzero $Q^2$.  

For photoproduction, the BS3 model reveals a different resonance structure mainly at forward and backward kaon angles that can be attributed to the energy-dependent widths and modification of the background part of the amplitude due to the choice of different hyperon resonances and re-fitting the parameters. 
A notable change is the disappearance of the second peak in the cross 
sections in the forward-angle region.

Predictions of the BS3 model at larger values of $Q^2$ [2--3 $(\text{GeV}/c)^2$], which are more sensitive to a choice of the electromagnetic form factors, are still not satisfactory. 
This together with our experience from fitting the model 
parameters to data varying the form factors of the $K^*$ and 
$K_1$ mesons suggest the necessity of further investigation 
of model properties for larger values of $Q^2$ given mainly 
by the shape of the form factors and their parameters.

\section*{Acknowledgements}
The authors thank Lothar Tiator and Reinhard Schumacher for useful discussions. This work was supported by the Grant Agency of the Czech Republic under the Grant No.~P203/15/4301 and by the JSPS Grant No.~16H03995.
\appendix

\section{Exchanges of $N^*$ in longitudinal coupling with virtual photon}
In the next sections, we summarize the invariant and scalar amplitudes for exchanges of nucleon resonances with spin 1/2, 3/2, and 5/2 in longitudinal coupling with a virtual photon. The invariant amplitudes are written with no hadron or electromagnetic form factor as form factors can be easily included by multiplying the coupling parameter with the given form factor.

\subsection{$N^*(1/2^{\pm})$ exchange in longitudinal coupling}
The vertex function for the electromagnetic $\gamma p N^*(1/2)$ vertex is given in Eq.~(\ref{eq:V-EM-1/2}). The strong vertex function reads
\begin{equation}
V_S = ig_S\Gamma_{\pm},
\end{equation}
and the amplitude for this contribution has the form 
\begin{equation}
\mathbb{M}_{LC}^{N^*(1/2)} = \bar{u}(p_\Lambda) V_S \frac{\not\! q+m_R}{s-m_R^2+i m_R \Gamma_R}V^{EM} u(p).
\end{equation}
The scalar amplitudes (see Ref.~\cite{BS} for definitions and more datails) are
 \begin{subequations}
 \begin{align}
 \mathcal{A}_1 = & \,\, \pm k^2\frac{G_3}{s-m_R^2+im_R\Gamma_R}, \\
 \mathcal{A}_2 = & \,\, \pm 2k^2\frac{G_3}{s-m_R^2+im_R\Gamma_R} = 2\mathcal{A}_1,\\
 \mathcal{A}_6 = & \,\, (m_R \mp m_p)\frac{-G_3}{s-m_R^2+im_R\Gamma_R},
 \end{align}
 \end{subequations}
where $G_3=g_S g_{3}^{EM}/(m_R+m_p)^2$ and the upper (lower) sign corresponds with the case of positive (negative) parity of the nucleon resonance.

\subsection{$N^*(3/2^{\pm})$ exchange in longitudinal coupling}
The vertex function for the electromagnetic $\gamma p N^*(3/2)$ vertex is given in Eq.~(\ref{eq:V-EM-3/2}) and the strong vertex function has the form
\begin{equation}
V^S_\nu = i \frac{g_S}{m_R m_K} \epsilon_{\mu\nu\lambda\rho}\Gamma_{\mp}\gamma_5\gamma^\lambda q^\mu p_K^\rho.
\end{equation}
With this knowledge we can write the amplitude for $N^*(3/2)$ exchange in the longitudinal coupling as follows
\begin{eqnarray}
\mathbb{M}_{LC}^{N^*(3/2)} = \bar{u}(p_\Lambda) V^S_\nu \frac{\not\! q + m_R}{s-m_R^2+i m_R \Gamma_R}\nonumber \\ \,\,\, \times \left(g^{\nu\beta}-\tfrac{1}{3}\gamma^\nu \gamma^\beta\right) V^{EM}_\beta u(p).
\end{eqnarray}
The scalar amplitudes are
 \begin{subequations}
 \begin{align}
 \mathcal{A}^\prime_1 ={}& -\tfrac{1}{3} G_3 k^2\left[-m_R (q\cdot p_\Lambda) \mp m_\Lambda s \right],\\
 \begin{split}
 \mathcal{A}^\prime_2 ={}& -\tfrac{1}{3} G_3 k^2 [2 m_R (q\cdot p_\Lambda) + m_R m_\Lambda m_p \\
 & \mp m_\Lambda s \mp 2 m_p(q\cdot p_\Lambda)], 
 \end{split} \\
 \mathcal{A}^\prime_3 ={}& -G_3 s k^2 (-m_R \pm m_p),\\
 \mathcal{A}^\prime_4 ={}& -\tfrac{1}{3} G_3 k^2 [-m_R m_\Lambda \pm 2(q\cdot p_\Lambda)], \\
 \mathcal{A}^\prime_5 ={}& \pm G_3 s k^2 ,\\
 \begin{split}
 \mathcal{A}^\prime_6 ={}& -\tfrac{1}{3} G_3 [-m_R m_p (q\cdot p_\Lambda) + m_R m_\Lambda s\\
 & - m_Rm_\Lambda (p\cdot k) \mp m_\Lambda m_p s \mp 2 (q\cdot p_\Lambda)s\\
 & \pm 2 (q\cdot p_\Lambda)(k\cdot p) \pm 3(p\cdot p_\Lambda)s],
 \end{split}
 \end{align}
 \end{subequations}
where $G_3=g_S g_{3}^{EM} / m_K m_R^2 (m_R+m_p)^2$ and the upper (lower) sign corresponds with the case of positive (negative) parity of the nucleon resonance. 

Each amplitude $\mathcal{A}_i, i=1,\ldots,6,$ has to be multiplied by the propagator denominator
\begin{equation}
\mathcal{A}_i = \frac{1}{s-m_R^2+im_R\Gamma_R} \mathcal{A}^\prime_i.
\end{equation}

\subsection{$N^*(5/2^{\pm})$ exchange in longitudinal coupling}
As the vertex function for the electromagnetic $\gamma p N^*(5/2)$ vertex is given in Eq.~(\ref{eq:V-EM-5/2}) and the vertex function for the interaction in the $K\Lambda N^*(5/2)$ vertex reads
\begin{equation}
V_S^{\mu\nu} = i \frac{g_S}{m_K^4} \Gamma_{\pm} q^2 p_K^\mu p_K^\nu,
\end{equation}
the amplitude for this contribution has the form
\begin{eqnarray}
\mathbb{M}_{LC}^{N^*(5/2)} = \bar{u}(p_\Lambda) V_S^{\mu\nu} \frac{\not\! q +m_R}{s-m_R^2 + im_R\Gamma_R}\mathcal{P}_{\mu\nu,\lambda\rho}^{(5/2)}(q) \nonumber \\ \!\! \times V_{EM}^{\lambda\rho} u(p),\ \ \  
\end{eqnarray}
where the spin-5/2 projection operator $\mathcal{P}_{\mu\nu,\lambda\rho}^{(5/2)}(q)$ reads
\begin{equation}
\begin{split}
\mathcal{P}_{\mu\nu;\lambda\rho}^{(5/2)}(q)={}&\tfrac{1}{2}(\mathcal{P}_{\mu\lambda}\mathcal{P}_{\nu\rho}+\mathcal{P}_{\mu\rho}\mathcal{P}_{\nu\lambda})-\tfrac{1}{5}\mathcal{P}_{\mu\nu}\mathcal{P}_{\lambda\rho}\\
& -\tfrac{1}{10}(\not\!\mathcal{P}_{\mu}\not\!\mathcal{P}_{\lambda}\mathcal{P}_{\nu\rho}+\not\!\mathcal{P}_{\mu}\not\!\mathcal{P}_{\rho}\mathcal{P}_{\nu\lambda}\\
& +\not\!\mathcal{P}_{\nu}\not\!\mathcal{P}_{\lambda}\mathcal{P}_{\mu\rho}+\not\!\mathcal{P}_{\nu}\not\!\mathcal{P}_{\rho}\mathcal{P}_{\mu\lambda}).
\end{split}
\end{equation}
The scalar amplitudes are
\begin{subequations}
\begin{align}
\mathcal{A}^\prime_1 = {}& -\tfrac{1}{5}G_3 k^2 A [\pm s\,m_\Lambda - m_R (q\cdot p_\Lambda)],\\
\begin{split}
\mathcal{A}^\prime_2 = {}& - \tfrac{1}{5} G_3 k^2 \{ 5 A(q\cdot p_\Lambda)(\pm m_p+m_R)-(\pm m_p + m_R) \\
& \times [s\,m_\Lambda^2 - (q\cdot p_\Lambda)^2][(q\cdot p)-s] \\
& +(m_\Lambda m_p m_R \pm sm_\Lambda) A - B(q\cdot p_\Lambda)(\pm m_p - m_R)\\
& - A(q\cdot p_\Lambda)(\pm m_p + m_R) \\
& - (q\cdot p_\Lambda)(\pm s -m_p m_R)C \} ,
\end{split}\\
\begin{split}
\mathcal{A}^\prime_3 = {}& \tfrac{1}{5} G_3 k^2 [5A\,s(\pm m_p + m_R)\\
& -B\,s(\pm m_p-m_R)-Cs(\pm s - m_R m_p)] ,
\end{split} \\
\begin{split}
\mathcal{A}^\prime_4 = {}& \tfrac{1}{5} G_3 k^2 \{ \pm 5A(q\cdot p_\Lambda) \pm [s m_\Lambda^2-(q\cdot p_\Lambda)^2][s-(q\cdot p)]\\
& \mp (q\cdot p_\Lambda)(A+B) + m_R [C(q\cdot p_\Lambda) + m_\Lambda A] \} ,
\end{split}\\
\mathcal{A}^\prime_5 = {}& -\tfrac{1}{5} G_3 sk^2[\pm 5A \mp B + m_R C],\\
\begin{split}
\mathcal{A}^\prime_6 = {}& -\tfrac{1}{5} G_3 \{ \mp 5 [(q\cdot p_\Lambda)(k\cdot p)-s(p_\Lambda\cdot k)]A \\
& \mp [s\,m_\Lambda^2 - (q\cdot p_\Lambda)^2][s-(q\cdot p)](k\cdot p)\\
& - (\pm B - m_R C)[s(k\cdot p_\Lambda)-(q\cdot p_\Lambda)(k\cdot p)] \\
& \mp (q\cdot p_\Lambda)[s-(k\cdot p)]A+m_\Lambda m_R [s-(k\cdot p)]A \\
& - [m_R(q\cdot p_\Lambda)\mp sm_\Lambda]m_p A\},
\end{split}
\end{align}
\end{subequations}
where $G_3=g_S g_{3}^{EM} / m_K^4 (2m_p)^5$, the upper (lower) sign corresponds with the case of positive (negative) parity of the nucleon resonance and 
\begin{subequations}
\begin{align}
A = {}& s(p_\Lambda \cdot p)-(q\cdot p_\Lambda) (q\cdot p), \\
B = {}& s\,m_\Lambda m_p -(q\cdot p_\Lambda)(q\cdot p),\\
C = {}& (q\cdot p_\Lambda)m_p-(q\cdot p)m_\Lambda.
\end{align}
\end{subequations}

Each amplitude $\mathcal{A}_i, i=1,\ldots,6,$ has to be multiplied by the propagator denominator
\begin{equation}
\mathcal{A}_i = \frac{1}{s-m_R^2+im_R\Gamma_R} \mathcal{A}^\prime_i.
\end{equation}

\end{document}